\documentclass{WileyMSP-template}
\usepackage{booktabs,chemformula}
\usepackage{color,soul}
\usepackage{pdflscape}
\usepackage{ragged2e}
\begin{document}
\justifying 
\pagestyle{fancy}
\rhead{\includegraphics[width=3cm]{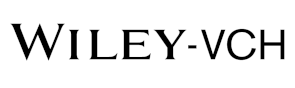}}
\setlength{\parindent}{1em}
\setlength{\parskip}{0.75em}

\title{\noindent Unveiling Phonon Contributions to Thermal Transport and the Failure of the Wiedemann-Franz Law in Ruthenium and Tungsten Thin Films}%

\maketitle

\author{Md. Rafiqul Islam$^1$},
\author{Pravin Karna$^2$},
\author{Niraj Bhatt$^2$},
\author{Sandip Thakur$^2$},
\author{Helge Heinrich$^3$},
\author{Daniel M. Hirt$^1$}
\author{Saman Zare$^1$},
\author{Christopher Jezewski$^4$},
\author{Rinus T.P. Lee$^{5}$},
\author{Kandabara Tapily$^{5}$},
\author{John T. Gaskins$^{6}$},
\author{Colin D. Landon$^{7}$},
\author{Sean W. King$^{8}$},
\author{Ashutosh Giri$^{2}$},
\author{Patrick E. Hopkins$^{1,9,10}$*}

\begin{affiliations}
\noindent$^1${Department of Mechanical and Aerospace Engineering, University of Virginia, Charlottesville, Virginia 22904, USA}\\
\noindent$^2$Department of Mechanical Industrial and Systems Engineering, University of Rhode Island, Kingston, RI 02881, USA\\
\noindent$^3$Nanoscale Materials Characterization Facility, University of Virginia, Charlottesville, Virginia 22904, USA\\
\noindent$^4$Technology Research, Intel Corporation, Hillsboro, Oregon 97124, USA\\
\noindent$^5$TEL Technology Center, America, Albany NY 12203, USA\\
\noindent$^6$Laser Thermal, Charlottesville, Virginia 22902, USA\\
\noindent$^7$Logic Technology Development, Intel Corporation, Hillsboro, Oregon 97124, USA\\
\noindent$^8$Supplier, Technology, and Industry Development, Intel Corporation, Hillsboro, Oregon 97124, USA\\
\noindent$^9${Department of Materials Science and Engineering, University of Virginia, Charlottesville, Virginia 22904, USA}\\
\noindent$^{10}$Department of Physics, University of Virginia, Charlottesville, Virginia 22904, USA

\noindent
Email Address: phopkins@virginia.edu
\end{affiliations}


\begin{abstract}

Thermal transport in nanoscale interconnects is dominated by intricate electron–phonon interactions and microstructural influences. As copper faces limitations at the nanoscale, tungsten and ruthenium have emerged as promising alternatives due to their substantial phonon contributions to thermal conductivity. Metals with stronger phonon-mediated thermal transport are particularly advantageous in nanoscale architectures, where phonons are less sensitive to size effects than electrons. Here, we show that phonons play a comparable role to electrons in the thermal transport of ruthenium and tungsten thin films, evidenced by deviations from the classical Wiedemann–Franz law. Elevated Lorenz numbers—1.9 and 2.7 times the Sommerfeld value for ruthenium and tungsten, respectively—indicate phonon contributions of 45\% and 62\% to total thermal conductivity. Comparisons of in-plane thermal conductivity from steady-state thermoreflectance and electron relaxation times from infrared ellipsometry reveal that phonon-mediated transport is insensitive to microstructural variations and scaling. Ultrafast infrared pump–probe measurements show that ruthenium exhibits a higher electron–phonon coupling factor than tungsten, consistent with the differing contributions of carriers to thermal transport. Molecular dynamics simulations and spectral energy density analysis confirm substantial phonon-driven thermal transport and mode-dependent phonon lifetimes. These results offer insights into phonon-driven thermal transport and provide design principles for selecting interconnects with enhanced thermal management.

\end{abstract}
\newpage
\section{Introduction}
Metallic nanowire interconnects are critical for linking transistors and ensuring efficient transport of signals and power. However, the relentless scaling of device dimensions introduces significant challenges, including increased RC delay and heat dissipation, which degrade interconnect performance. \cite{islam2024evaluating,perez2022dominant,gall2020search} Although copper (Cu) is the most commonly used interconnect, it faces challenges such as thermochemical and thermomechanical stability at the nanoscale, along with increased resistivity due to boundary scattering.\cite{liu2020accommodation,gall2020search} For example, when dimensions shrink to around 25 nm, \cite{islam2024evaluating} the resistivity of Cu can rise dramatically, complicating its viability for future electronic applications. The current technology node in very-large-scale integration (VLSI) is expected to shrink below 5~nm, driven by the growing demand for faster devices capable of meeting the increasing performance and energy efficiency requirements of modern electronic systems.\cite{zhan2020effect} To address these challenges, researchers are exploring alternative materials that may offer superior performance characteristics. Two promising candidates for interconnects are tungsten (W) and ruthenium (Ru), which exhibit lower electromigration enhancing reliability and competitive electrical resistivity compared to Cu.\cite{croes2018interconnect,nogami2017comparison,gall2016electron,bhatt2023transition,teng2012reliability} Ruthenium ($\sim$6.0 nm) and tungsten ($\sim$15.5 nm) possess shorter electron mean free paths compared to Cu ($\sim$39 nm), \cite{gall2016electron,karna2024electron} indicating potential advantages in the reduced sensitivity to boundary scattering effects on electronic charge and thermal transport relative to Cu.\cite{gall2020search} Unlike Cu, both W and Ru also exhibit substantial phonon contributions to thermal conductivity. However, the influence of these contributions on thermal transport at reduced sizes remains uncertain, with studies reporting contradictory or inconclusive findings regarding electron and phonon size effects in Ru and W. \cite{wang2024size,bhatt2023transition,hu2021first,stojanovic2010thermal} These contributions challenge the traditional Wiedemann-Franz law, which correlates thermal and electrical conductivities based on electron transport.\cite{bhatt2023transition,chen2019understanding} Such findings emphasize the need for further investigation into nanoscale thermal transport mechanisms in Ru and W, particularly to account for the significant role of phonons in these materials.

Despite significant advancements in computational frameworks for predicting the contributions of phonons and electrons to the thermal conductivity of  W and Ru single crystals, a substantial gap remains in understanding how morphology—such as grain size, grain boundary segregation, or defects influence phonon behavior in metals.\cite{stojanovic2010thermal,wang2024size} The complexity of accurately modeling phonon-phonon and electron-phonon interactions further complicates this field. Consequently, the thermal transport properties of these metals remain inadequately understood, particularly regarding how characteristics of reduced grain sizes and layer thicknesses affects thermal conductivity, thus contributing to uncertainties surrounding the Wiedemann-Franz law. This knowledge gap hinders the search for advanced interconnects capable of meeting the demands of modern electronics.  Experimental methods have the potential to address these limitations by measuring the contributions of phonons and electrons in W and Ru while incorporating the effects of thickness and grain morphology—an area that is still largely underdeveloped.

In this study, we synthesize Ru and W films of various thickness using physical vapor deposition (PVD). We employ Scanning Transmission Electron Microscopy (STEM) to analyze the films' morphology, crystal structure, and chemical composition. Our focus is on the thermal conductivity and electron-phonon coupling factor of  Ru and W thin films to investigate the effects of thickness and grain morphology on thermal transport. We employ a steady-state thermoreflectance (SSTR) technique to measure their in-plane thermal conductivity. By comparing our experimental measurements with values derived from electrical resistivity, we assess the applicability of the Wiedemann-Franz law, particularly using the low-temperature value of the Lorenz number ($L_0$) to calculate thermal conductivity. Our findings reveal significant deviations from the Wiedemann-Franz law when evaluating the total thermal conductivity of Ru and W. Notably, we observe that the Lorenz number for these thin films consistently exceeds the Sommerfeld value ($L_0$), by $1.9 \times L_0$ and $2.7 \times L_0$ for Ru and W films, respectively. This deviation can be attributed to phonon contributions, which account for approximately $45\%$ (Ru) to $62\%$ (W) of the total thermal conductivity. The measured phonon thermal conductivity of Ru and W aligns well with values determined using a machine-learned potential (MLP) developed from \textit{ab initio} molecular dynamics (AIMD) simulations. Importantly, the phonon contribution is found to be independent of grain size and thickness. To gain deeper insights into phonon thermal conductivity across varying film thicknesses and grain size, we utilize an ultrafast (sub-picosecond) pump-probe technique. This method allows us to monitor the intraband transient thermoreflectance response of Ru and W films, enabling us to extract the electron-phonon coupling factor ($G$). Our results indicate that $G$ is higher in Ru films than in W films, indicating that electron-phonon scattering plays a dominant role in determining phonon lifetimes in Ru. This observation aligns well with mode-dependent phonon lifetimes obtained through spectral energy density calculations and our first-principles calculations based on the full electron-phonon coupling matrix. Our study successfully determines the phonon contribution to thermal conductivity in Ru and W films across different thicknesses and grain sizes, validates the Lorenz number deviations, and provides a comprehensive understanding of the interplay between thickness, grain morphology, and phonon scattering mechanisms.

\begin{figure}[htb]
\begin{center}
\includegraphics[width=\textwidth]{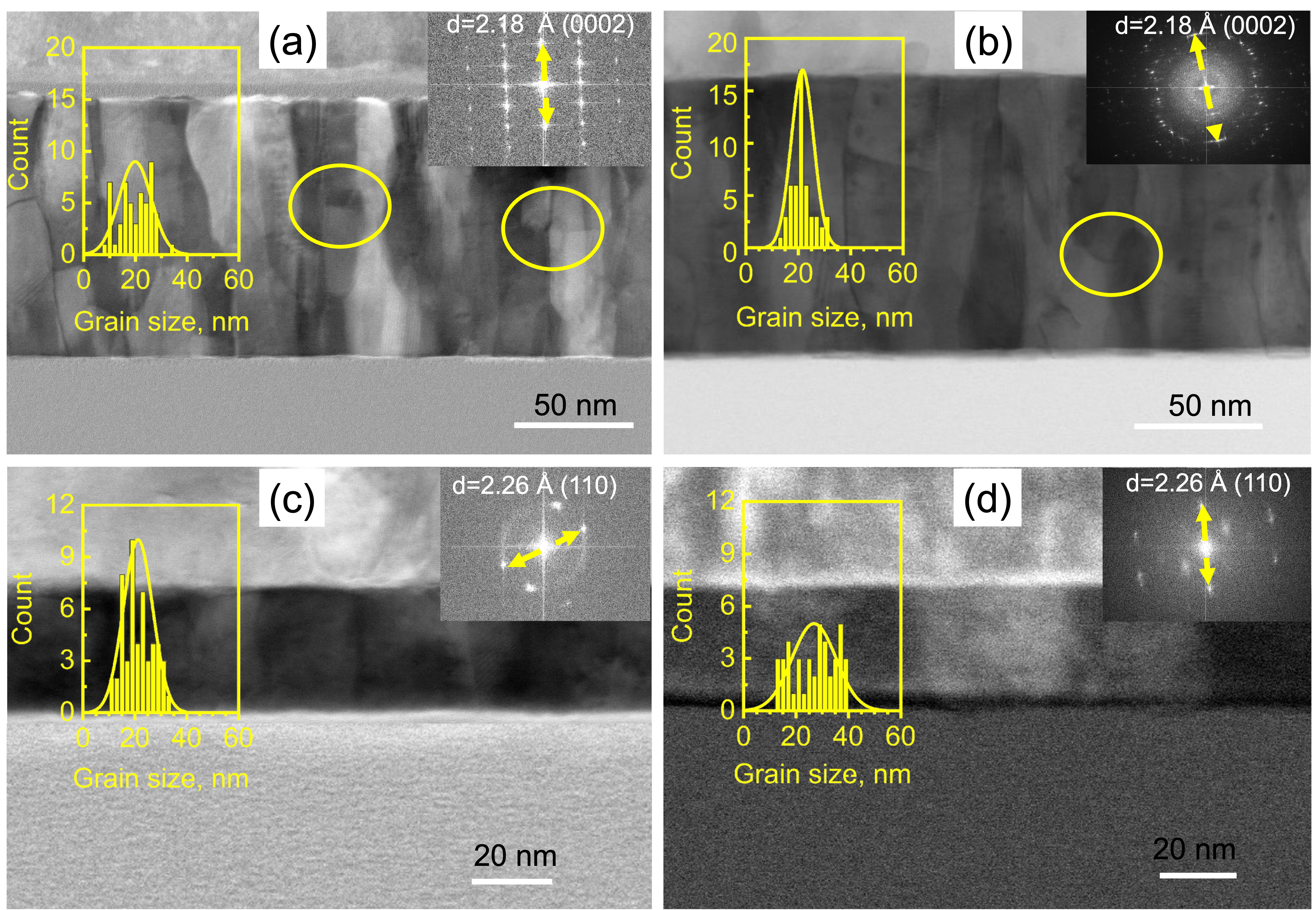}
\caption{STEM image of (a), as-deposited $\approx$ 102 nm Ru, (b) 400 \textdegree C annealed $\approx$ 102 nm Ru films, (c) as-deposited $\approx$ 30 nm W\(_{\text{R}}\) and (d) as-deposited $\approx$ 31 nm W\(_{\text{G}}\), respectively. The micrographs show the representative area of the as-deposited films. The grain size distribution of the films is shown in the inset. Ru exhibits a mixed grain morphology where the dominant structure is columnar, but some grains exhibit irregular or equiaxed forms (marked by circle). Annealing does not eliminate the mixed grain morphology in the Ru films, but it increases the proportion of larger grains. The grains of W films are mostly columnar. The FFT patterns obtained from selected regions of the high-resolution STEM of these as-deposited films confirm the HCP structure of Ru and and the BCC structure of W.
\label{FigTEM}}
\end{center}
\end{figure}

\section{Results and Discussion}

We investigate the influence of microstructure and thickness on the thermal transport properties of Ru and W films by analyzing their growth, grain morphology, and crystallographic characteristics. Using the physical vapor deposition (PVD) method, we synthesize a series of Ru films with thicknesses ranging from 5 nm to 100 nm on 100 nm SiO$_2$/Si substrates. Additionally, we deposit W films with thicknesses ranging from 3 nm to 101 nm on 3 nm Ta/100 nm SiO$_2$/Si substrates using the PVD technique. The W films are synthesized using two distinct approaches: one designed to reduce resistivity, referred to as W$_{\text{R}}$, and the other aimed at modifying grain size, labeled as W$_{\text{G}}$. Both film types are deposited in a 300 mm multi-cathode PVD chamber. For W$_{\text{R}}$, the deposition utilizes lower cathode power and reduced ionization, which results in smaller grain sizes. We characterize the microstructure, chemical composition, and crystallography of selected Ru and W films using STEM.  Figure~\ref{FigTEM}a presents the mixed grain morphology of an as-deposited $\approx$102 nm Ru film, where columnar grains dominate, while some grains exhibit irregular or equiaxed forms (marked by circles). Certain grains transition from an equiaxed structure near the interface to a columnar structure at the top as they grow vertically. This variation likely results from inhomogeneities during the growth process, such as fluctuations in deposition rate, substrate surface roughness, or local temperature variations, which influence nucleation and growth dynamics. \cite{pan2010effect, behura2017chemical, wolf1990inhomogeneous} The grain size distribution of this $\approx$102 nm Ru film appears in the inset of Fig.~\ref{FigTEM}a. Similar mixed grain morphologies are observed in Ru films with thicknesses between 20 nm and 50 nm (detailed in the Supporting Information), with the effect becoming more pronounced as the film thickness increases. To refine the Ru film microstructure, we anneal a subset at 400 $^{\circ}$C for one hour in an argon atmosphere. Although annealing does not eliminate the mixed grain morphology in the $\approx$102 nm Ru film, it increases the proportion of larger grains, as shown in Fig.~\ref{FigTEM}b and its inset. Table~1 lists the grain sizes of both as-deposited and annealed Ru films. The persistence of mixed grain morphology, even after annealing, suggests that grain growth is constrained by existing boundaries and structural irregularities. 

In contrast to Ru films, W films exhibit a predominantly columnar grain morphology. Figures~\ref{FigTEM}c-d show the columnar structure of the $\approx$30 nm W$_{\text{R}}$ and $\approx$31 nm W$_{\text{G}}$ films, respectively. The grain size of W$_{\text{G}}$ is slightly larger than that of W$_{\text{R}}$, as indicated in the insets of Figs.~\ref{FigTEM}c-d and detailed in Table~1. The grain size distribution further reveals that thicker W$_{\text{G}}$ films generally exhibit larger grains than their thinner counterparts (see Supporting Information). These variations in grain morphology significantly influence thermal transport in Ru and W films. The mixed grain morphology in Ru increases grain boundary density and introduces additional interfaces that enhance phonon and electron scattering, leading to reduced thermal conductivity. In contrast, the columnar grain structure in W films minimizes grain boundary scattering, facilitating more efficient heat transport. A more detailed analysis of these effects is provided in the following section. We confirm the structural integrity of the Ru and W films through high-resolution STEM imaging and fast Fourier transform (FFT) pattern analysis. Ru films crystallize into a hexagonal close-packed structure, while W films adopt a body-centered cubic structure. We further detect no impurities when we performed compositional analysis of the Ru and W films using STEM with energy-dispersive X-ray spectroscopy (EDX). Additionally, the low surface roughness of the Ru films ($\approx$1 nm) minimizes light scattering during pump-probe experiments, ensuring that the measured signals reflect the intrinsic material properties rather than surface-induced artifacts (further details on chemical compositions and surface roughness are available in the Supporting Information).

\captionsetup[table]{labelformat=simple, labelsep=colon} 
\setlength{\tabcolsep}{.3em} 

\begin{table*}[h!]
\caption{\label{tab:films} Thickness and average lateral grain size of the films determined from STEM images over a large area. The thickness of PVD films is found to be uniform (see Supporting Information for additional details).}
\centering
\resizebox{\textwidth}{!}{ 
\begin{tabular}{@{}p{2.5cm} p{2.5cm} p{2.5cm} p{2.5cm} p{2.5cm} p{2.5cm} p{2.5cm} p{2.5cm}@{}}
\toprule
\textbf{Thickness of} & \textbf{Grain size of} & \textbf{Thickness of} & \textbf{Grain size of} & \textbf{Thickness of} & \textbf{Grain size of} & \textbf{Thickness of} & \textbf{Grain size of} \\
\textbf{as-deposited Ru film (nm)} & \textbf{as-deposited Ru films (nm)} & \textbf{annealed Ru (nm)} & \textbf{annealed Ru (nm)} & \textbf{as-deposited W\(_{\text{R}}\) (nm)} & \textbf{as-deposited W\(_{\text{R}}\)(nm)} & \textbf{W\(_{\text{G}}\) (nm)} & \textbf{W\(_{\text{G}}\) (nm)} \\
\midrule
\large $\sim$ 10 & \large 6.7 $\pm$ 1.1 &  &  & \large $\sim$ 22 & \large 16.6 $\pm$ 3.1 &  &  \\
\large $\sim$ 22 & \large 16.0 $\pm$ 4.6 &  &  & \large $\sim$ 26 & \large 20.6 $\pm$ 4.6 &  &  \\
\large $\sim$ 32 & \large 21.9 $\pm$ 3.2 &  &  & \large $\sim$ 30 & \large 21.2 $\pm$ 5.6 & \large $\sim$ 31 & \large 26.8 $\pm$ 8.3 \\
\large $\sim$ 52 & \large 21.2 $\pm$ 3.2 &  &  &  &  & \large $\sim$ 102 & \large 38.2 $\pm$ 9.8 \\
\large $\sim$ 102 & \large 19.8 $\pm$ 6.2 & \large $\sim$ 102 & \large 21.6 $\pm$ 4.3 &  &  &  &  \\
\bottomrule
\end{tabular}
}
\end{table*}

\begin{figure}
\begin{center}
\includegraphics[width=\textwidth]{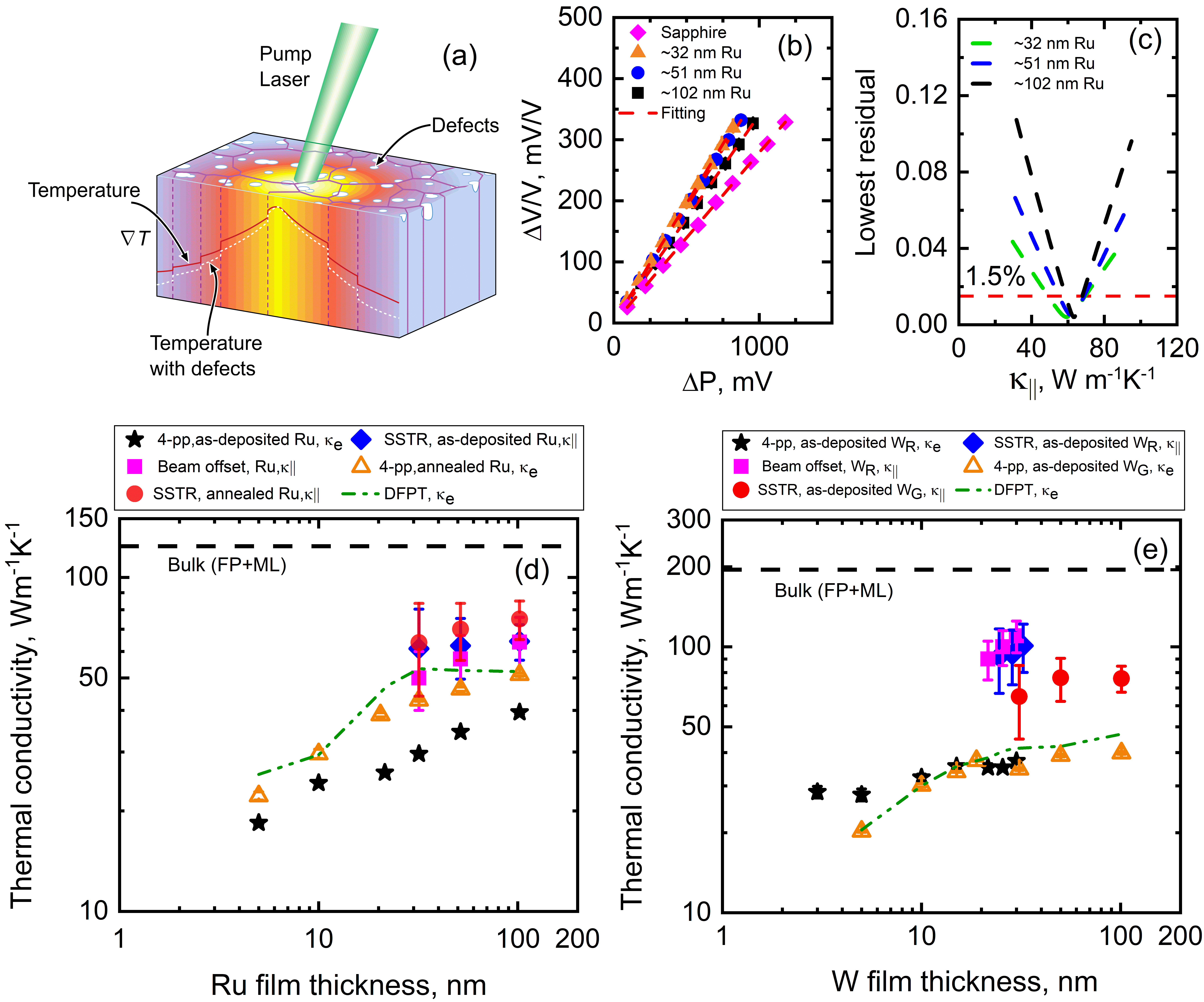}
\caption{(a) Schematic illustrating in-plane heat spreading during an SSTR experiment. Grain boundaries and defects scatter electrons and phonons, impeding heat flow and leading to a higher temperature rise near the pump laser spot and a steeper temperature gradient (\(\nabla T\)) compared to an ideal defect-free material. The dashed line represents the modified temperature profile caused by defects. 
(b) Plot of \(\frac{\Delta V}{V}\) vs. \(\Delta P\) (where \(\Delta P \propto\) pump power) used in the SSTR fitting to extract the thermal conductivities of Ru films. 
(c) 2D contour analysis of Ru films with different thicknesses, assuming the quality of fit between model parameters and experimental data is within a 1.5\% threshold value. 
(d) and (e) In-plane thermal conductivity (\(k_{\parallel}\)) of the as-deposited Ru and W films measured with SSTR, shown by blue diamond symbols. The SSTR-measured in-plane thermal conductivity of the as-deposited Ru and W films is higher than that derived from the Wiedemann--Franz law applied to electrical resistivity measurements (indicated by the star symbol), assuming the low-temperature Lorenz number. This difference arises from the phonon contribution to the thermal conductivity of Ru (\(27.4 \pm 3.0\, \mathrm{W\, m^{-1}\, K^{-1}}\), determined using a machine-learned potential developed from \textit{ab initio} molecular dynamics simulations) and W (\(\approx 60\, \mathrm{W\, m^{-1}\, K^{-1}}\)).\cite{bhatt2023transition} The short-dashed line indicates the predicted electron thermal conductivity of Ru and W thin films as a function of film thickness. The intrinsic bulk electron thermal conductivity, obtained from density functional perturbation theory (DFPT), is modified to account for boundary scattering (arising from finite film thickness and grain size), point defect scattering, and electron--phonon interactions through Matthiessen’s rule. While the model exhibits strong agreement with experimental data for W, it slightly overestimates the values for Ru, which is attributed to mixed grain morphology not fully captured by the current modeling framework.
\label{FigCuTC}}
\end{center}
\end{figure}

The thermal transport properties of Ru and W films are evaluated to explore the roles of thickness and grain morphology, as well as to determine phonon contributions, with a focus on in-plane thermal conductivity measured using the SSTR technique. SSTR  is a pump-probe method that operates at a low pump modulation frequency (e.g., 1000 Hz) to ensure that the pump heating induces steady-state thermal gradients. This steady-state regime enhances the sensitivity of SSTR to the in-plane thermal conductivity of thin films, making it more effective than other thermoreflectance techniques under similar experimental conditions (such as comparable pump and probe spot sizes, shapes, and alignment) .\cite{braun2019steady,hoque2021high,islam2024evaluating} Figure \ref{FigCuTC}a illustrates the in-plane heat spreading in a SSTR experiment. Grain boundaries and defects scatter electrons and phonons, and impede heat flow, resulting in a higher temperature rise near the pump laser spot and a steeper temperature gradient compared to an ideal defect free material, which SSTR can effectively capture due to its high sensitivity to in-plane thermal transport.\cite{hoque2021high,islam2024evaluating} 

To ensure consistent opto-thermal transduction across all Ru and W film thicknesses and to increase the sensitivity to their in-plane thermal conductivity these films are coated with a 20 nm Al/60 nm Ti transducer. This transducer has been effective in precisely measuring the in-plane thermal conductivity of metal films in our prior work,\cite{islam2024evaluating} further establishing its reliability for this study. To quantify the thermal conductivity of the films, we proceed with the SSTR fitting process, which involves analyzing the relationship between the measured voltage change and the power change.  Figure \ref{FigCuTC}b illustrates the relationship between $\Delta V/V$ and $\Delta P$, which we use in SSTR fitting to extract the thermal conductivities of sapphire and as-deposited Ru films. The detailed methodology is described elsewhere.\cite{braun2019steady,islam2024evaluating} Briefly, we determine a proportionality constant between $\Delta V/V$ and $\Delta P$ using a sapphire calibration sample with well-characterized thermal properties for both the transducer and the sapphire. The key assumption is that this proportionality constant remains unchanged across the samples of interest.  Once the proportionality constant is established from the sapphire, we determine the in-plane thermal conductivity of Al/Ti-coated Ru films using a steady-state thermal model. The analysis of Ru films is based on a five-layer model comprising the Al/Ti transducer, Ru thin film, SiO$_2$, and Si substrate. For the thin films, we observe considerable sensitivity to various thermal parameters. To address this, we employ contour plots to characterize the similarity between experimental data and thermal model fits as shown in Fig \ref{FigCuTC}c, enabling us to resolve the sensitivity by measuring the in-plane thermal conductivity of Ru, indicated by the narrowness of the curve. In this technique, our measured SSTR data are compared to a thermal model, iterating the in-plane thermal conductivity of Ru to achieve the best fit. Figure \ref{FigCuTC}c demonstrates the quality of the fit, with a residual within 1.5\%, highlighting the capability of SSTR to accurately measure in-plane thermal conductivity of film thickness $\geq$ 30 nm.  The residual error between the theoretical model and experimental data reaches a maximum of approximately 1.5\%, indicating that multiple combinations of thermal conductivity and boundary resistance can achieve a similar fit. Based on this, we set a 1.5\% residual threshold, ensuring that the model consistently produces high-quality fits to the experimental data. \cite{feser2012probing,islam2024evaluating} This type of contour residual analysis represents a specific source of uncertainty: the uncertainty associated with fitting the thermal model to the experimental data. Note that the uncertainty increases as the thickness decreases, and for Ru films $\leq 30$ nm, we lose sensitivity to their in-plane thermal conductivity. In the case of W\(_{\text{R}}\), we can measure in-plane thermal conductivity down to $\approx 22$ nm, while for W\(_{\text{G}}\), we maintain sensitivity down to $\approx 30$ nm. Details on the extraction of thermal conductivity for W films and the associated contour uncertainties are provided in the Supporting Information. Additional details on the SSTR experimental setup, analysis, and uncertainties are also provided in the Methods section and Supporting Information.

Figures \ref{FigCuTC}d-e present the in-plane thermal conductivities (\( k_{\mid\mid} \)) of as-deposited Ru, W\(_{\text{R}}\) and W\(_{\text{G}}\) films, measured using the SSTR technique. A key observation is that the in-plane thermal conductivity of the as-deposited Ru films, represented by blue diamond symbols, exhibits no discernible dependence on thickness. This behavior is primarily attributed to the grain size distribution. As shown in Table 1, the average grain size remains consistent across all measured Ru films, indicating that the grain boundary density does not vary significantly with thickness. Consequently, the structural uniformity across different thicknesses explains the absence of a pronounced thickness dependence in thermal conductivity. Moreover, although \( \approx \) 102 nm Ru film has a thickness much larger than its electron-phonon mean free path, its thermal conductivity remains lower than the bulk value, \cite{ho1972thermal} highlighting the impact of microstructural features on thermal transport. This reduced thermal conductivity, relative to bulk Ru, is attributed to the films’ mixed grain morphology, where non-columnar grains coexist with columnar ones. This microstructural heterogeneity introduces a higher density of grain boundaries and misaligned domains, which act as scattering centers for electrons. The inset of Figures 1a–b, along with the Supporting Information, reveals that a portion of the grains is comparable to or smaller than the electron mean free path. In regions with a high density of grain boundaries, increased electron scattering further impedes thermal transport, thereby increasing thermal resistance and leading to a lower thermal conductivity than bulk Ru.  


 For W films, the in-plane thermal conductivity decreases relative to its bulk value and remains relatively constant within the measurement uncertainty, exhibiting behavior similar to that of Ru films. Although these W films do not exhibit mixed grain morphology, they possess a uniformly columnar structure, which minimizes grain boundary scattering and facilitates more efficient heat transport in the cross-plane direction. Moreover, the average grain size of W\(_{\text{R}}\) films is comparable to and approaches the electron mean free path length of W (\( \lambda_{\mathrm{ep}} \approx 15.5 \, \mathrm{nm} \)),~\cite{karna2024electron,gall2016electron} thereby contributing to a reduction in thermal conductivity. Similarly, while the grain size of W\(_{\text{G}}\) films is larger than that of W\(_{\text{R}}\), their in-plane thermal conductivity remains comparable within experimental uncertainty. This observation is attributed to the thermal boundary resistance (\(5.0 \, \mathrm{m^2 \, K \, GW^{-1}}\))
 \cite{islam2024evaluating, hoque2021thermal} assumed between tungsten and SiO\(_2\), which is considered consistent across Ru, W\(_{\text{R}}\), and W\(_{\text{G}}\) films. The precise determination of this boundary resistance is beyond the scope of this study.  

We validate the accuracy of our SSTR measurements for Ru and W films by comparing them with results obtained using the beam-offset pump-probe technique. This method, recognized for its precision in measuring in-plane thermal conductivity, leverages lateral heat flow induced by the pump.~\cite{qian2020accurate,feser2012probing} By strategically offsetting the pump and probe beams, this technique isolates in-plane thermal transport while minimizing the influence of the 100 nm SiO\(_2\)/Si substrate, which could otherwise introduce measurement artifacts. The thermal conductivity values obtained from both the beam-offset and SSTR techniques exhibit excellent agreement, falling within the respective uncertainty ranges. This consistency further reinforces the reliability of our measurements and underscores the robustness of our experimental approach.

To assess the applicability of the Wiedemann-Franz law in determining the in-plane thermal conductivity of Ru and W, we compare our steady-state thermoreflectance (SSTR) measurements with values derived from electrical resistivity (shown as star symbols). The Wiedemann-Franz law establishes a proportionality between the electrical and thermal conductivities of metals, assuming that conduction is primarily governed by free electrons and that the ratio of thermal to electrical conductivity is constant, determined by the Lorenz number (\( L_0 = 2.44 \times 10^{-8} \, \mathrm{W} \, \Omega \, \mathrm{K}^{-2} \)) in the Sommerfeld theory.\cite{ashcroft2022solid,ziman2001electrons,white1960lorenz} This principle is widely applicable in bulk metals under equilibrium conditions at low temperatures, where lattice vibrations have minimal contributions. However, in our thin-film systems, significant deviations are observed between the SSTR and 4-point probe measured in-plane thermal conductivities, highlighting the limitations of the Wiedemann-Franz law when using \( L_0 \) as the Lorenz number. 

Notably, the experimental Lorenz number for Ru and W\(_{\text{R}}\) films exceeds the Sommerfeld value by factors of \( 1.9 L_0 \) and \( 2.7 L_0 \), respectively. These deviations indicate that additional heat transport mechanisms—beyond purely electronic contributions—play a significant role in these films. While an increased experimental Lorenz number might initially suggest a modification to \( L_0 \) is necessary to describe electron-mediated thermal transport, this interpretation does not fully capture the underlying physics. Instead, the observed deviations arise due to the substantial phonon contribution to thermal conductivity, rather than a fundamental breakdown of the Wiedemann–Franz law in predicting electronic thermal conductivity. Since the total thermal conductivity is given by \( k_{\text{total}} = k_e + k_p \), where \( k_e \) is derived from electrical conductivity using \( L_0 \), the excess thermal transport must be attributed to phononic contributions. For Ru, the phonon thermal conductivity is measured as \( 25.5 \pm 2.7\, \text{W}\,\text{m}^{-1}\,\text{K}^{-1} \) for film thicknesses \( \geq 30 \) nm. Similarly, for W films (thicknesses \( \geq 20 \) nm), the phonon thermal conductivity is measured as \( 47.1 \pm 12.5\, \text{W}\,\text{m}^{-1}\,\text{K}^{-1} \). These values indicate that the phonon contribution accounts for a substantial portion of the total thermal conductivity, even in materials traditionally considered dominated by electronic heat transport.

The measured phonon thermal conductivity aligns well with our theoretical predictions, reinforcing that the standard \( L_0 \) remains valid for predicting the electronic component (\( k_e \)), while the additional thermal conductivity originates from phonon-mediated transport. We predict the thermal conductivity of Ru using a machine-learned potential (MLP) developed from \textit{ab initio} molecular dynamics (AIMD) simulations. This computational approach accurately captures the lattice dynamics and phonon interactions, showing strong agreement with \textit{ab initio} results. Using the Green-Kubo formalism and equilibrium molecular dynamics (EMD), the phonon thermal conductivity of Ru at room temperature is calculated from the heat current autocorrelation function (HCACF) as \( 27 \pm 3 \, \mathrm{W} \, \mathrm{m}^{-1} \, \mathrm{K}^{-1} \), with uncertainties below \( 12\% \) (details are provided in the Supporting Information). The measured phonon thermal conductivity of W is consistent with our previous work.~\cite{bhatt2023transition} The enhanced lattice contribution in W is attributed to weak anharmonic phonon scattering, driven by relatively weak interatomic interactions that limit three-phonon processes. In contrast, the larger scattering phase space and stronger anharmonicity in Ru lead to more frequent phonon–phonon interactions, which reduce phonon lifetimes and suppress lattice thermal conductivity. The role of phonon–phonon and electron–phonon interactions on phonon lifetime is discussed in detail in the following sections. Taken together, these results confirm that the elevated Lorenz numbers stem from phonon-mediated transport, not from deviations in electronic behavior. Thus, the standard Sommerfeld value \( L_0 \) remains valid for estimating the electronic thermal conductivity \( \kappa_e \), with the excess thermal transport accurately attributed to the lattice contribution. 

\begin{figure}[htb]
\begin{center}
\includegraphics[width=\textwidth]{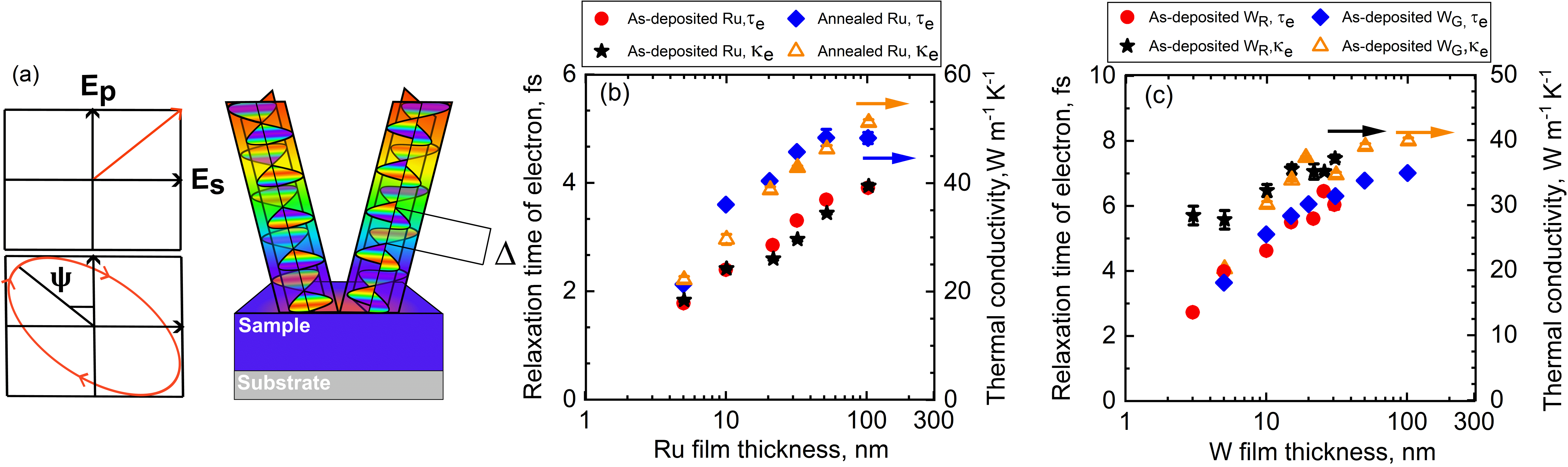}
\caption{(a) Schematic representation of spectroscopic ellipsometry for thin film characterization.(b) The effective relaxation time of as-deposited Ru (represented by red circle symbols) determined from IR-VASE measurements exhibits a thickness-dependent trend that aligns with the behavior of electron thermal conductivity. For annealed Ru films, the relaxation time increases due to the larger length scale of electron-grain boundary interaction. 
(c) Similarly, the effective relaxation time for as-deposited W\(_{\text{R}}\) and pure W\(_{\text{G}}\) films follows a size-dependent trend, consistent with data obtained from 4-point-probe measurements.
\label{FigACuTC}}
\end{center}
\end{figure}

Our findings suggest that the failure of the Wiedemann-Franz law to accurately predict the thermal conductivity of W and Ru arises from the substantial phonon contribution to the total thermal conductivity. Additionally, size effects in electrical and thermal transport within polycrystalline films are more pronounced in electron transport than in phonon transport. These results further indicate that applying the low-temperature value of the Lorenz number, \( L_0 \), is appropriate for predicting the electronic contribution to the total thermal conductivity in W and Ru. To further validate our hypothesis that the Wiedemann-Franz law, when used with \( L_0 \), reliably predicts electron thermal conductivity, we employ density functional perturbation theory (DFPT) calculations to predict the electron thermal transport properties of Ru and W thin films. The intrinsic electron thermal conductivity is first calculated, and then boundary scattering—arising from film thickness and grain size—as well as point defect scattering, are incorporated using Matthiessen’s rule. Additionally, electron–phonon scattering rates obtained from DFPT are included to ensure a comprehensive treatment of transport mechanisms. The DFPT-predicted electron thermal conductivity (represented by the dashed line) shows good agreement with values measured via the 4-point probe method (star and triangle symbols) for W films. In the case of as-deposited Ru films, the DFPT predictions slightly overestimate the experimental values. This discrepancy is attributed to the presence of mixed grain morphology, where grain sizes are comparable to or smaller than the electron mean free path, as revealed by the grain size distributions (Fig.~\ref{FigTEM}) and summarized in Table~1.
 While grain size effects are included in the modeling, the DFPT-based approach does not fully capture the influence of mixed grain morphology on electron thermal conductivity.
\begin{figure}
\begin{center}
\includegraphics[width=\textwidth]{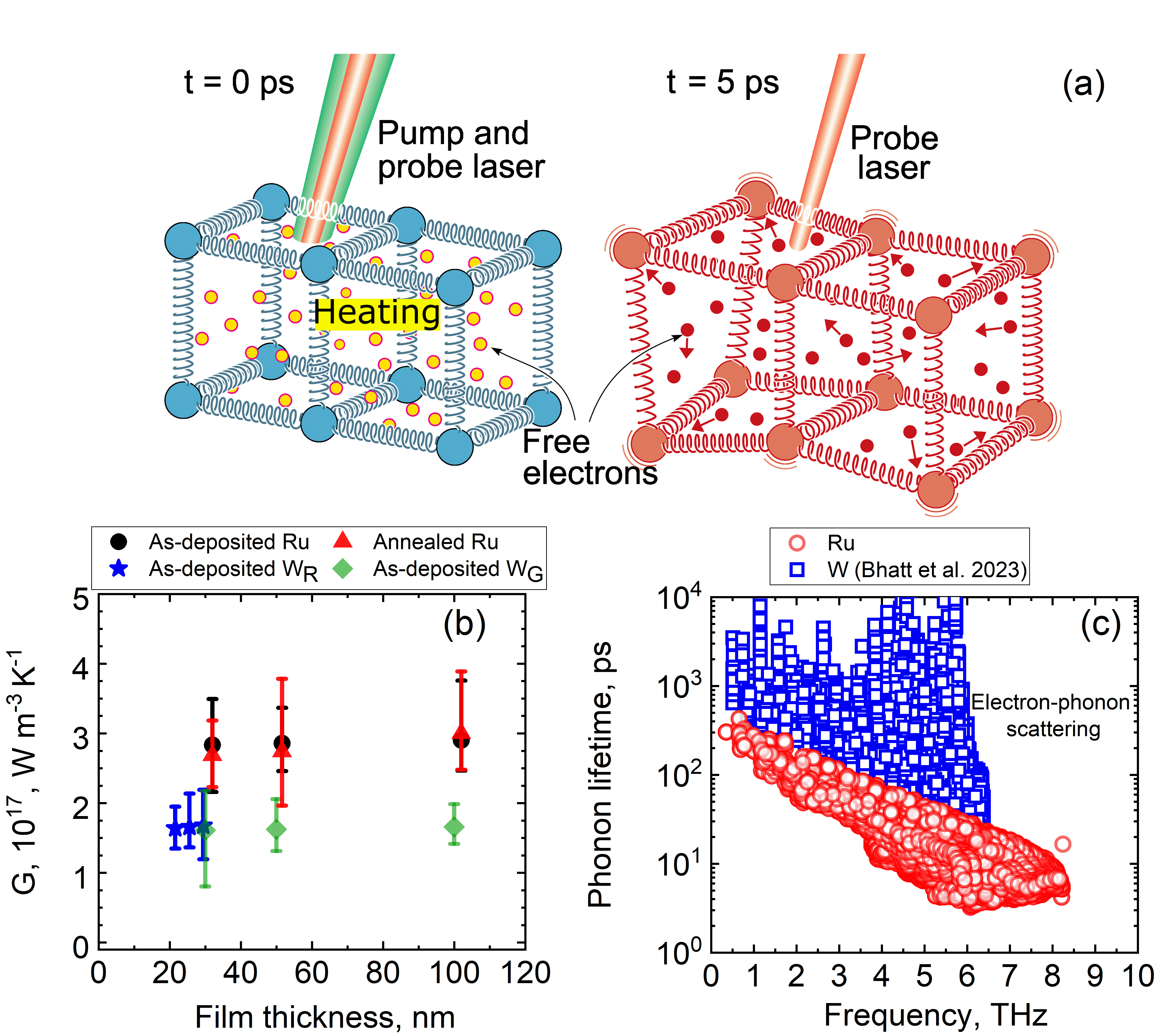}
\caption{(a) The electron-phonon scattering mechanism under the laser excitation. (b) Two temperature model derived electron–phonon coupling factor ($G$) of the as-deposited films. The measured coupling factor of Ru is higher than that of W, indicating that more electron-phonon scattering events occur in Ru, which results in shorter phonon lifetime in Ru as found from our first-principles calculations as shown in (c).
\label{FigRuWP}}
\end{center}
\end{figure}

\begin{figure}[htb]
\begin{center}
\includegraphics[width=\textwidth]{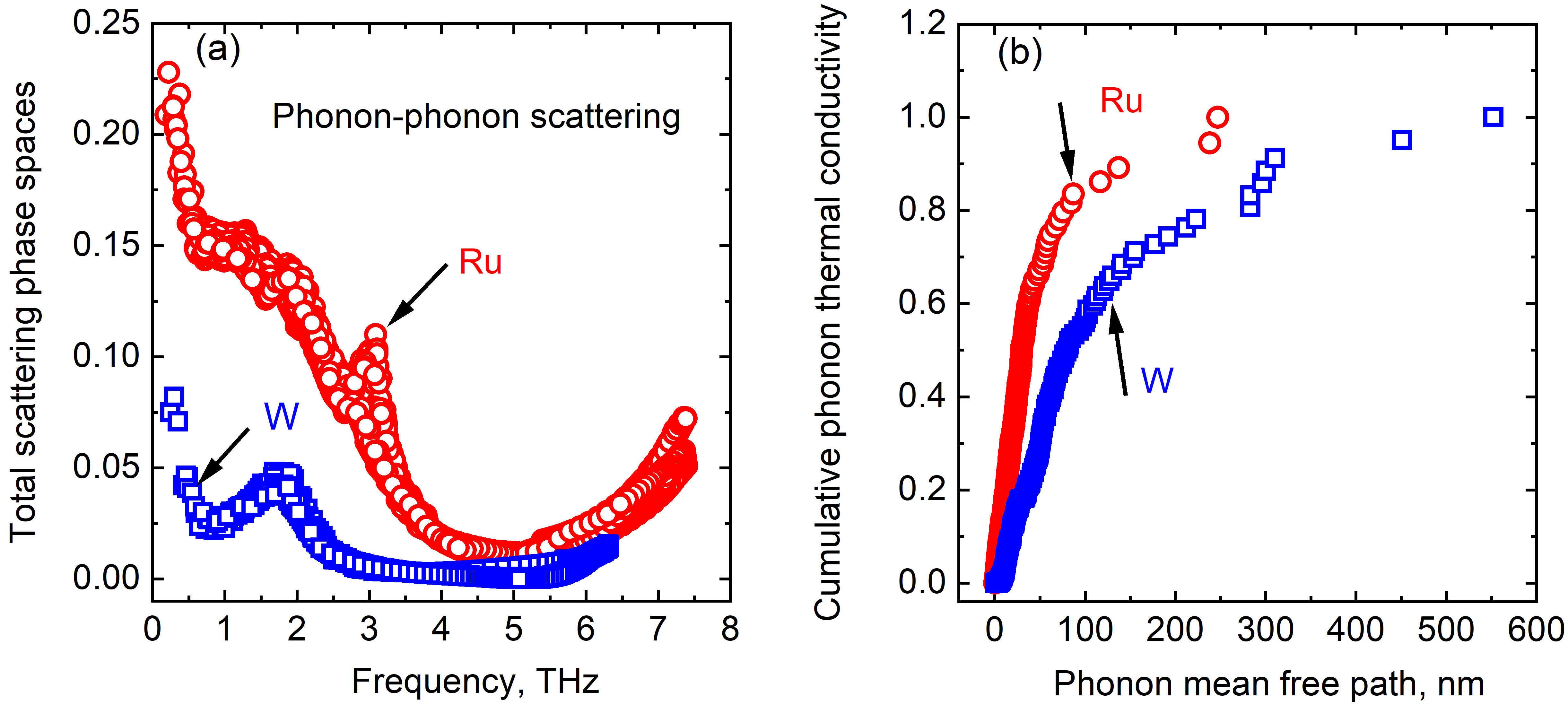}
\caption{(a) Total phonon-phonon scattering phase space (SPS) as a function of phonon frequency for Ru (red circles) and W (blue squares). The significantly larger SPS in Ru indicates higher phonon-phonon scattering rates, contributing to reduced phonon lifetimes and lower thermal conductivity compared to W. (b)Normalized cumulative phonon thermal conductivity as a function of phonon mean free path for Ru (red circles) and W (blue squares). In Ru, a substantial fraction of the phonon thermal conductivity arises from phonons with mean free paths below $\sim$50~nm, whereas in W, phonons with longer mean free paths contribute more significantly.
\label{FigRuWEP}}
\end{center}
\end{figure}

The size-dependent reduction in the thermal conductivity of thin metal films is primarily attributed to increased electronic scattering rates. To further support this, we measure electron scattering rates using infrared variable-angle spectroscopic ellipsometry (IR-VASE). Figure~\ref{FigACuTC}a illustrates the principle of spectroscopic ellipsometry used for characterizing thin-film samples. Upon interaction with the sample surface, the incident polarized light undergoes a change in polarization, resulting in an elliptically polarized reflected beam. These polarization changes, analyzed through optical modeling, yield insights into the optical properties and electron relaxation times in Ru and W films. The electronic relaxation times measured via IR-VASE show a strong correlation with electron thermal conductivity, providing a direct link between electron scattering and thermal transport. This approach bridges the gap between theoretical predictions and experimental observations, highlighting the interplay between electron scattering mechanisms and thermal transport properties. Figure~\ref{FigACuTC}b shows that the relaxation time in as-deposited Ru films is thickness-dependent, decreasing with increasing film thickness. This trend mirrors the behavior observed in electron thermal conductivity, further supporting the role of boundary scattering in limiting electron transport. Annealed Ru films exhibit longer relaxation times compared to their as-deposited counterparts, which can be attributed to grain growth during annealing. In contrast, phonon transport in Ru films appears to be independent of both film thickness and grain morphology, as evidenced by SSTR measurements, electron thermal conductivity data, and relaxation time analyses. The phonon thermal conductivity of Ru films with thicknesses $\geq 30$\,nm is measured to be $25.5 \pm 2.7$\,W\,m$^{-1}$\,K$^{-1}$, representing the average value across samples, with the uncertainty corresponding to the standard deviation. This consistency underscores the minimal impact of boundary scattering on phonon transport in Ru thin films. Indeed, cumulative phonon thermal conductivity analysis  (see Fig.~\ref{FigRuWEP}b) reveals that a substantial portion of thermal conductivity in Ru is contributed by phonons with  mean free path below $\sim$50~nm, particularly in the range $\sim$30--50~nm, where boundary scattering becomes less influential. We discuss boundary scattering effects on phonon transport in more detail in the following section, based on molecular dynamics simulations using the spectral energy density formalism.  These findings align with prior theoretical studies, which suggest that grain boundary resistance significantly influences electron transport but has minimal effect on phonon transport.~\cite{stojanovic2010thermal,hu2021first} Similarly, the relaxation times extracted for W films also exhibit a thickness dependence, as shown in Fig.~\ref{FigACuTC}c. However, this trend is less pronounced compared to that observed in Ru. The phonon thermal conductivity of W films remains constant across varying thicknesses and grain sizes, as the majority of heat is carried by phonons with mean free path well below $\sim$100~nm. The independence of phonon transport from film thickness and grain morphology in both Ru and W contrasts with materials such as Cu, where phonon contributions are minimal, and electrons experience enhanced resistivity due to boundary scattering. Collectively, these observations provide a comprehensive understanding of how grain morphology and boundary scattering strongly influence electron transport in Ru, while having a weaker or negligible effect in W. In both materials, phonon transport remains largely unaffected, laying the groundwork for future investigations into other transport-limiting mechanisms.

The phonon contribution to thermal conductivity in metals is primarily governed by electron-phonon and phonon-phonon interactions. \cite{chen2019understanding, bhatt2023transition, wang2016first, tong2019comprehensive, jain2016thermal}The pronounced differences in phonon thermal conductivity between Ru and W arise from the weaker phonon-phonon and electron-phonon interactions in W, which reduce phonon scattering and enhance phonon transport. To explore this phenomenon further and gain deeper insights into the phonon thermal conductivity of Ru and W thin films, we measure the electron-phonon coupling factor and uncover critical details about the spectral-level characteristics of electron-phonon interactions, including mode-dependent phonon lifetimes limited by electron-phonon scattering. These measurements provide a comprehensive understanding of how electron-phonon coupling impacts thermal transport mechanisms in these films.

To measure the electron-phonon interactions, we utilize a sub-picosecond pump (2.38 eV)-probe (0.539–0.56 eV) technique to study Ru and W films with thicknesses ranging from 30 nm to 102 nm for Ru and from 20 nm to 100 nm for W. Figure~\ref{FigRuWP}a illustrates the electron scattering phenomena induced by laser excitation.  The near-infrared probe energy is particularly advantageous for monitoring the ultrafast dynamics of excited electrons in these metals, as it lies significantly below the interband transition energies of W ($0.94 \pm 0.05$, $1.80 \pm 0.015$ eV) and Ru ($1.74 \pm 0.02$ eV, $2.79\pm 0.01$), which are determined using visible and infrared VASE (details in the supporting information). By employing this probe energy, we capture a nearly free-electron-like thermoreflectance response, where the resulting temporal dynamics predominantly reflect electron scattering near the Fermi surface. These dynamics provide direct insight into diffusive electronic transport while minimizing interference from interband transitions. Additionally, probing the free-electron dynamics after pump heating also simplifies the extraction of the electron–phonon coupling factor ($G$) from the transient optical response of the films. This parameter is crucial for quantifying energy transfer rates between the electron system and the lattice. Our approach enables precise determination of $G$, as previously reported in studies. \cite{islam2024evaluating,tomko2021temperature} These measurements not only reveal the coupling strength but also shed light on the mode-specific influence of electron-phonon interactions on phonon lifetimes and thermal conductivity. Together, they provide a unified framework to analyze the interplay between electronic and lattice contributions to heat transport in Ru and W thin films.

Figure~\ref{FigRuWP}b presents the extracted  $G$ from ultrafast thermoreflectance measurements, performed at probe photon energies of approximately 0.53 eV and 0.56 eV, using a two-temperature model (TTM; details provided in the Method section and Supporting Information). The coupling factor remains relatively unchanged between the as-deposited and annealed Ru films, suggesting that electron–phonon scattering is not significantly influenced by variations in grain size or film thickness. We also derive $G$ for W films of different thicknesses and find the values to be consistent. The derived $G = 2.9 \times 10^{17} \, \text{Wm}^{-3}\text{K}^{-1}$ for Ru is higher than the value derived for W films, $G = 1.7 \times 10^{17} \, \text{Wm}^{-3}\text{K}^{-1}$. This difference arises due to the higher electron density of states around the Fermi surface and the presence of optical phonon modes in Ru.~\cite{karna2024electron, bhatt2023transition} The high electron density of states within the Fermi window provides more available electron states for electron-phonon scattering, which leads to a higher coupling factor. However, our measured $G = 2.9 \times 10^{17} \, \text{Wm}^{-3}\text{K}^{-1}$ for Ru is approximately 9 times lower than our first-principles calculations and approximately 3.5 times lower than previously reported experimental values.~\cite{bonn2000ultrafast,jang2020nonequilibrium} The measured coupling factor aligns with some other works based on first-principles calculations,~\cite{medvedev2020electron,milov2018modeling,milov2018mechanism} highlighting a wide discrepancy in reported values.  This  discrepancy may arise from the sensitivity of the TTM fitting to several input parameters, particularly the electron heat capacity and thermal conductivity, both of which depend strongly on the electronic density of states and are temperature-dependent. While we have determined the thermal conductivity using SSTR and 4-point probe, and other parameters using IR-VASE measurements and calculations, the electron heat capacity is not measured experimentally. Uncertainties in their temperature dependence, along with possible simplifications in the TTM framework, could contribute to the underestimation of $G$. Moreover, the relatively large upper-bound uncertainty in the TTM fit suggests that the actual electron–phonon coupling in Ru may be higher.  In contrast, the derived value of $G = 1.7 \times 10^{17} , \text{Wm}^{-3}\text{K}^{-1}$ for W agrees well with previous first-principles calculations for films of similar thicknesses.~\cite{lin2008electron,karna2024electron} This consistency indicates that the assumptions regarding the electron heat capacity coefficient and other modelinputs at elevated electron temperatures are reasonable in the case of W. Overall, despite the discrepancy in absolute values, the ultrafast thermoreflectance data clearly indicate a higher electron–phonon coupling in Ru compared to W, consistent with previous reports in the literature. The details of the modeling parameters and ultrafast thermoreflectance measurements are provided in the Supporting Information.

The higher electron-phonon coupling factor for Ru indicates that energy transfer between electrons and the lattice occurs more rapidly in Ru than in W. This rapid energy exchange leads to shorter phonon lifetimes in Ru films compared to W films, as phonons are more frequently scattered by interactions with electrons. Consequently, the reduced phonon lifetimes in Ru diminish the phonon contribution to its overall thermal conductivity, shifting the balance of heat transport toward electronic conduction. This trend is corroborated by our SSTR and 4-point probe measurements, which reveal a lower lattice-driven thermal contribution in Ru compared to W. We  also calculate the phonon lifetimes using first-principles methods based on the full electron-phonon coupling matrix to examine the impact of the electron-phonon coupling factor on phonon lifetimes. Figure~\ref{FigRuWP}c shows that Ru phonon lifetimes are shorter than those of W at all frequencies due to higher electron-phonon scattering rates. The presence of optical modes further reduces the phonon lifetimes, leading to lower phonon thermal conductivity for Ru.

We further validate phonon lifetimes and phonon thermal conductivity by investigating phonon-phonon interactions through Spectral Energy Density (SED) calculations. We employ SED analysis to obtain the total scattering phase space (SPS), presented in Fig.~\ref{FigRuWEP}a, quantifying the number of allowed three-phonon scattering processes as a function of frequency. Additional details on phonon absorption, including Normal (N) and Umklapp (U) processes, as well as phonon emission mechanisms, are provided in the Supporting Information. Ru exhibits a significantly larger SPS across the phonon frequency spectrum, particularly in the low-to-mid frequency range, highlighting its greater propensity for phonon-phonon scattering. Conversely, W demonstrates a lower SPS, indicating fewer available scattering channels and reduced phonon-phonon interactions.
These findings emphasize intrinsic differences in the bulk phonon transport properties of Ru and W: Ru experiences stronger phonon scattering and lower lattice thermal conductivity, whereas W retains higher intrinsic phonon transport due to reduced scattering phase space. Despite these differences, both materials exhibit thickness-independent phonon thermal conductivity at the nanoscale, though at different characteristic length scales. As shown in Fig.\ref{FigRuWEP}b, a substantial portion of phonon thermal conductivity in Ru arises from phonons with mean free paths below $\sim$50nm, rendering boundary scattering effects minimal in films with thicknesses $\geq$30nm. In contrast, the phonon thermal conductivity of W remains constant even down to $\sim$20nm films, indicating that most heat-carrying phonons have even shorter MFPs and are similarly unaffected by boundary scattering within this thickness range.

\section{Summary}
Our work unveils the significant role of phonon-mediated thermal transport in nanoscale ruthenium and tungsten thin films, with phonons contributing up to 62\% of the total thermal conductivity. The observed deviation from the classical Wiedemann–Franz law, coupled with microstructure-insensitive phonon transport and distinct electron–phonon coupling behaviors, highlights a paradigm shift in understanding heat conduction in metals at nanoscale. Supported by molecular dynamics simulations and spectral energy density analysis, these insights offer a new framework for selecting and engineering interconnect materials optimized for thermal performance in advanced nanoelectronic architectures.

\section{Experimental and Computational Methods}

\textbf{Morphology and chemical composition analysis.}  We examine the grain structure and chemical composition of Ru and W thin films using STEM. We prepare cross-sectional samples with a Thermo Fisher Helios system and thin the cross-section to below 50~nm. To reduce damage from the gallium ion beam during milling, we deposit a protective platinum layer on the film surface. We perform STEM analysis using a Thermo Fisher Themis Z-STEM operated at 200~kV and equipped with a SuperX detector for compositional analysis. We acquire both bright-field STEM and high-angle annular dark-field (HAADF) images simultaneously; bright-field images help us measure grain sizes, while HAADF images provide atomic number contrast. Using the SuperXG2 EDX detector in the STEM mode, we carry out simultaneous imaging and elemental mapping without switching modes. The samples are aligned in an edge-on orientation by observing the diffraction pattern of the Si substrate. To quantify grain size distributions, we use ImageJ software for statistical analysis. We include this analysis along with representative STEM images of Ru and W films with different thicknesses.

\noindent
\textbf{In-plane thermal conductivity measurements.} 
We use the steady-state thermoreflectance (SSTR) technique to measure the in-plane thermal conductivity of our samples. This method relies on the linear correlation between surface temperature rise and reflectivity change.~\cite{braun2019steady,hoque2021high} We utilize a 637~nm continuous-wave pump laser, modulated at 1000~Hz, onto the sample surface to generate steady heating. A 785~nm probe laser monitors the reflectivity change resulting from this temperature increase. Both laser beams travel through optical fibers and converge at a common 20$\times$ microscope objective for precise alignment.~\cite{foley2021fiber} We detect the probe signal variation at the modulation frequency using a lock-in amplifier, which measures the amplitude of the reflected probe light. By adjusting the pump power and keeping the modulation frequency constant, we observe changes in the probe response. This response scales linearly with the input power and corresponds to the ratio between temperature rise and heat flux, characterized by a proportionality factor $\gamma$. To determine $\gamma$, we measure a reference material—typically sapphire—using an identical transducer layer. We assume that $\gamma$ remains constant between the reference and target samples. After obtaining $\gamma$ from the reference, we extract the thermal conductivity of unknown samples using a steady-state heat conduction model.~\cite{braun2019steady} Further details of the procedure are available in the Supporting Information.

\noindent
\textbf{First principles calculations}.
We calculate the electron thermal conductivity and electron-phonon coupling factor ($G$) of Ru and W thin films using first-principles methods based on Density Functional Perturbation Theory (DFPT) and the Electron-Phonon Wannier (EPW) framework within \texttt{Quantum ESPRESSO}.\cite{giannozzi2009quantu,cai2015giant} A detailed interpolation of electron-phonon matrix elements from coarse to fine \textit{k}- and \textit{q}-grids is performed using maximally localized Wannier functions (MLWFs).~\cite{marzari2012maximally} Norm-conserving pseudopotentials from the PS Library are used for both elements.~\cite{troullier1991efficient} The Eliashberg spectral function $\alpha^2F(\omega)$ and mass enhancement parameter $\lambda$ are computed, yielding values of $\lambda \approx 0.52$ for Ru and $\approx 0.29$ for W, consistent with previous reports.~\cite{karna2024electron} The average electron mean free paths are found to be approximately 6 nm for Ru and 15.5 nm for W, based on the calculated electron-phonon scattering rates. The corresponding Fermi velocities are approximately $6 \times 10^5$ m/s for Ru and $9.4 \times 10^5$ m/s for W. At room temperature, the calculated values of $G$ are approximately $2 \times 10^{17}$ W·m$^{-3}$·K$^{-1}$ for W and $2.5 \times 10^{18}$ W·m$^{-3}$·K$^{-1}$ for Ru. The resulting electronic thermal conductivities are 135 W m$^{-1}$ K$^{-1}$ for W and 97 W·m$^{-1}$·K$^{-1}$ for Ru.

To account for boundary scattering and structural effects in thin films, we model the total electron scattering rate using:~\cite{hopkins2011boundary,he2023hydride}
\begin{equation}
\frac{1}{\tau_{\text{tot}}} = \frac{v_F}{\Lambda} + \frac{v_F}{d} + \frac{v_F}{D} + \frac{v_F}{\Lambda_{\text{defect}}}
\end{equation}
where $v_F$ is the Fermi velocity, $\Lambda$ is the mean free path due to electron-phonon scattering, $d$ is the film thickness, $D$ is the grain size, and $\Lambda_{\text{defect}} = \frac{a}{\sqrt{c}}$ represents the mean free path associated with point defect scattering, where $a$ is the lattice constant and $c$ is the defect concentration.~\cite{klemens1955scattering} In this study, we assume a point defect concentration of 0.2\% in both Ru and W thin films.~\cite{lu2017using,nakagawa1979spontaneous} Electron-electron scattering is neglected, as electron-phonon interactions dominate at room temperature.

\noindent
\textbf{Computational evaluation of thermal conductivity and phonon dynamics}. 
We develop a machine-learned interatomic potential (MLP) for ruthenium using a diverse dataset generated from \textit{ab-initio} molecular dynamics (AIMD) simulations performed in \texttt{Quantum ESPRESSO}.~\cite{giannozzi2009quantum} We sample atomic configurations under various pressures and lattice distortions at ambient temperatures, collecting 5,000 training frames and 1,000 validation frames containing atomic positions, energies, forces, and cell parameters. Using the DEEPMD-kit~\cite{zhang2018deep} integrated with TensorFlow,~\cite{developers2022tensorflow} we train the MLP through a deep neural network consisting of a ResNet-like embedding network and a fitting network, and we optimize the model with the Adam algorithm and a decaying learning rate. The validation results show excellent agreement with DFT data, yielding energy and force root mean square errors (RMSEs) of $8.13 \times 10^{-3}$ eV/atom and $46.87 \times 10^{-3}$ eV/\AA, respectively. We perform molecular dynamics simulations using LAMMPS~\cite{thompson2022lammps} and the trained MLP to calculate the lattice thermal conductivity based on the Green-Kubo formalism. We used the equation:
\begin{equation}
\kappa = \frac{1}{Vk_B T^2} \int_0^\infty \langle \mathbf{J}(0) \cdot \mathbf{J}(t) \rangle \, dt
\end{equation}

where $V$ is the volume, $T$ the temperature, $k_B$ the Boltzmann constant, and $\mathbf{J}(t)$ the heat current at time $t$. We equilibrate the system under NPT, NVT, and NVE ensembles, then ran 10 ns production simulations with a 10 fs sampling interval to compute the heat current autocorrelation function. By averaging five independent simulations at room temperature, we estimate the lattice thermal conductivity of Ru to be $27.4 \pm 3$ W/m$\cdot$K, with uncertainties under 12\%, demonstrating the robustness and accuracy of our MLP.

We perform molecular dynamics simulations using the spectral energy density (SED) formalism to investigate the atomic-scale dynamics of the material. This method involves applying a Fourier transform to atomic velocities to compute the average kinetic energy as a function of wave vector ($\mathbf{q}$) and frequency ($\omega$). The SED expression accounts for atomic mass, velocity components, unit cell positions, and total simulation time, and has been described in detail in prior work.~\cite{feng2015anharmonicity} The expression is given by:

\begin{align}
\Phi(\mathbf{q}, \omega) = \frac{1}{4\pi \tau N_T} \sum_{\alpha=1}^{3} \sum_{b=1}^{B} m_b \left| \int_0^{\tau} \sum_{n_{x,y,z}}^{N_T} \dot{u}_\alpha(\mathbf{n}_{x,y,z}, b; t) \, e^{i \mathbf{q} \cdot \mathbf{r}(\mathbf{n}_{x,y,z}, 0) - i \omega t} \, dt \right|^2
\end{align}

Here, $\tau$ is the total simulation time, $\alpha$ denotes the Cartesian direction, $b$ is the atomic index in a unit cell, $B$ is the number of atoms in the unit cell, and $m_b$ is the mass of atom $b$. $\dot{u}_\alpha$ represents the velocity component in the $\alpha$-direction, and $\mathbf{r}(\mathbf{n}_{x,y,z}, 0)$ is the equilibrium position of each unit cell $\mathbf{n}_{x,y,z}$. $N_T$ is the total number of unit cells in the supercell. For the calculations, we construct a $4 \times 100 \times 4$ supercell and equilibrate the system using the NPT ensemble for 2 ns with a timestep of 0.5 fs.~\cite{hoover1985canonical} We follow this with a 2 ns simulation under the NVT ensemble. Finally, we perform a 1.5 ns NVE simulation, during which we record atomic velocity data for SED analysis.

\noindent
\textbf{Electron-phonon interactions measurements}. 
We investigate electron scattering mechanisms in Ru and W thin films using an ultrafast pump-probe technique with sub-picosecond temporal resolution and tunable infrared probe wavelengths to capture the intraband transient thermoreflectance response. A Nd:YVO$_4$ laser operating at a 1 MHz repetition rate and centered at $\sim$1040 nm provides the primary beam, which is directed through an optical parametric amplifier (OPA). Within the OPA, we generate a 520 nm (2.38 eV) pump beam by focusing part of the laser through a second harmonic generation (SHG) crystal. This pump beam amplifies another portion of the laser that passes through variable delay stages and a second SHG crystal, producing a tunable probe beam spanning 600 nm (2.06 eV) to 2500 nm (0.5 eV). We mechanically chop the pump pulses (379 fs duration) at 450 Hz and determine their temporal width by fitting the thermoreflectance signals of Pt using a $\textnormal{sech}^2$ function.~\cite{islam2024evaluating} During measurements, we excite the electron system in the Ru and W films at 2.38 eV and probe their thermoreflectivity at $\sim$0.53 eV and $\sim$0.56 eV, respectively. These probe energies lie well below the interband transition thresholds of W ($0.94 \pm 0.05$ and $1.80 \pm 0.015$ eV) and Ru ($1.74 \pm 0.02$ and $2.79 \pm 0.01$ eV), allowing us to capture a predominantly free-electron-like response. We determine the interband transition energies using visible ellipsometry. To extract the electron-phonon coupling factors in both materials, we fit the thermoreflectivity data using the two-temperature model (TTM):

\begin{align}
C_e(T_e) \frac{\partial T_e}{\partial t} &= \nabla \cdot (\kappa_e \nabla T_e) - G(T_e - T_p) + S(x,t) \\
C_p(T_p) \frac{\partial T_p}{\partial t} &= \nabla \cdot (\kappa_p \nabla T_p) + G(T_e - T_p)
\end{align}
where $C_e$ and $C_p$ are the electron and phonon heat capacities, $T_e$ and $T_p$ are their respective temperatures, and $S(x,t)$ represents the energy source term.

\noindent
\textbf{Effective electron relaxation time determination}.We determine the interband transition energies of Ru and W thin films using a visible ellipsometer (M2000, J.A. Woollam Company) and an infrared ellipsometer (IR-VASE Mark II, J.A. Woollam Company). For Ru, we identify interband transitions at \(1.74 \pm 0.02\)~eV and \(2.79 \pm 0.01\)~eV, while for W, the corresponding transitions occur at \(0.94 \pm 0.05\)~eV and \(1.80 \pm 0.015\)~eV. We acquire the ellipsometric spectra using the IR ellipsometer in the spectral range of 500--3500~cm\(^{-1}\) (0.06--0.433~eV) at incident angles of 60\degree{} and 70\degree{}, with a spectral resolution of 16~cm\(^{-1}\) (approximately 2~meV). Supplementary Figure 23 presents the measured data for as-deposited Ru with thicknesses ranging from 5~nm to 102~nm. To extract the complex, frequency-dependent dielectric function, we construct an isotropic multi-layer optical model based on the measured spectra. In the infrared regime, the optical response of both Ru and W films is dominated by free-electron behavior and low-energy interband transitions. To account for these effects, we employ the Drude-Lorentz model:~\cite{fowles1989introduction, shi2013preparation}

\begin{equation}
\epsilon_{\text{Drude-Lorentz}}(\omega) = \epsilon_1(\omega) + i\epsilon_2(\omega) = \epsilon_\infty - \frac{\omega_p^2}{\omega^2 + i\Gamma\omega} + \sum_i \frac{f_i}{\omega_i^2 - \omega^2 - i\Gamma_i \omega}
\end{equation}

In this expression, \(\epsilon_\infty\) is the high-frequency dielectric constant. The second term represents the free-electron contribution, where \(\omega_p\) is the plasma frequency and \(\Gamma\) is the electron scattering rate. The final term accounts for interband electronic transitions, modeled using Lorentz oscillators,~\cite{fowles1989introduction} where \(f_i\), \(\omega_i\), and \(\Gamma_i\) denote the oscillator strength, resonance frequency, and broadening parameter of the \(i\)-th oscillator, respectively. We optimize all model parameters by minimizing the mean square error between the simulated and experimental ellipsometric data. From the fitted scattering rate \(\Gamma\), we calculate the effective electron relaxation time using \(\tau = \Gamma^{-1}\).

\medskip
\textbf{Supporting Information} \par
Supporting Information is available and has been provided with the submission.

\medskip
\textbf{Acknowledgements} \par 
We appreciate financial support from the Semiconductor Research Corporation, Grant No. 2025-NM-3286 and the National Science Foundation, Grant No. 2318576.

\medskip
\textbf{Conflict of interest} \par
The authors declare no conflict of interest.

\medskip
\textbf{Author contributions} \par
    M.R.I, P.E.H and C.D.L designed the experiments. C.J., S.W.K., R,T.P.L. and K.P. synthesized the films. M.R.I, D.M.H., P.K., S.Z., A.G., J.T.G., performed the experiments and analysis. P.K.,N.B.,S.T.,M.R.I and A.G. performed the computations. H.H., performed the TEM characterization. M.R.I. and P.E.H wrote the manuscript.   
    
\medskip
\textbf{Data availability statement} \par
    The data that support the findings of this study are available from the corresponding author upon reasonable request.

%
\bibliographystyle{MSP}
\bibliography{MegaRefs}

\end{document}